\documentclass[aps,floats,amssymb,twocolumn,prb]{revtex4}
\usepackage{calc}
\usepackage{psfrag}
\usepackage{graphicx}
\usepackage{color}
\usepackage{bm}

\begin{document}

\title{Pairing instabilities of Dirac composite fermions}

\author{M. V. Milovanovi\'c$^{1}$, M. Dimitrijevi\'c \'Ciri\'c$^{2}$, and V. Juri\v ci\'c$^{3}$}
\affiliation{$^{1}$ Scientific Computing Laboratory, Center for the Study of Complex Systems, Institute of Physics Belgrade, University of Belgrade, Pregrevica 118, 11080 Belgrade, Serbia}
\affiliation{$^{2}$ Faculty of Physics, University of Belgrade, 11001 Belgrade, Serbia}
\affiliation{$^{3}$ Nordita, Center for Quantum Materials, KTH Royal Institute of Technology
and Stockholm University, Roslagstullsbacken 23, 10691 Stockholm, Sweden}

\begin{abstract}
 Recently, a Dirac (particle-hole symmetric) description of composite fermions in the half-filled  Landau level (LL) was proposed [D. T. Son, Phys. Rev. X {\bf 5}, 031027 (2015)], and we study its possible consequences on BCS (Cooper) pairing of composite fermions (CF's). One of the main consequences is the existence of anisotropic states in single and bilayer systems, which was previously suggested in  Ref. [J. S. Jeong and K. Park, Phys. Rev. B {\bf 91}, 195119 (2015)]. We argue that in the half-filled LL in the single layer case the gapped states may sustain anisotropy, because isotropic pairings may coexist with anisotropic ones. Furthermore, anisotropic pairings with addition of a  particle-hole (PH) symmetry breaking  mass term may evolve into rotationally symmetric states, i.e. Pfaffian states of Halperin-Lee-Read (HLR) ordinary CF's.  On the basis of the Dirac formalism, we argue
that in the quantum Hall bilayer at total filling one, with decreasing distance between layers weak pairing of $p$-wave paired CF's is gradually transformed from Dirac to ordinary, HLR - like, with concomitant  decrease in the CF number. Global characterization of low-energy spectrum based on the Dirac CF's agrees well
with previous calculations performed by exact diagonalization on a torus.
Finally, we discuss features of Dirac formalism when applied in this context.
\end{abstract}

\maketitle
% ------------------------------------
\section{Introduction}
% ------------------------------------
Composite fermions (CF's) [\onlinecite{jain}] describe the physics of electrons in fractional quantum Hall (FQH) regime. At filling factor $\nu=1/2$, essentially they absorb the external flux, and make a metallic state [\onlinecite{hlr}] with its own Fermi surface - Fermi surface of CF's. By slightly modifying Read's dipole construction of composite (neutral) fermions in the half-filled lowest Landau level [\onlinecite{read}], an argument can be given for the accumulation of Berry phase equal to $\pi$ as a CF encircles its own Fermi surface [\onlinecite{ws}].
This has motivated a description of the CF's in this setting in terms of Dirac fermions, which has been recently introduced in Ref. [\onlinecite{son}], and have attracted some interest [\onlinecite{ws,gz,psv,msone,kach,mrf,btj,bmf,mv}].
The PH symmetric description of the half-filled LL is given in terms of a Dirac system of composite quasiparticles - Dirac CF's at a finite chemical potential [\onlinecite{ws}] and in the presence of a gauge field. However, the implied existence of singularity at zero momentum in the CF
spectrum was criticized [\onlinecite{fdmh,bj}].
We may add that, due to the requirement of gauge invariance in two dimensions (2D), a small mass
must be introduced into the Dirac theory (``parity anomaly") [\onlinecite{red}]. This may be a way
to heal and complete in the high-energy domain (``UV completion" [\onlinecite{ws2}]) Dirac
description of CF's, and avoid singularity.

Thus the description in terms of Dirac fermions
may have capacity to capture essential, at least qualitative, aspects of the CF's physics. To further examine this possibility in this work we consider BCS pairing of Dirac CF's. First, in the framework of Dirac description of a single CF, we point out that, assuming Cooper pairing between spinor components, besides so-called PH symmetric Pfaffian, also anisotropic states  can be realized. This is  analogous to the ${^3}$He system in which both  $B$ and $A$ (anisotropic) phases are possible [\onlinecite{vol}].  Next we discuss unconventional $p$-wave  pairing of two kinds of Dirac CF's, motivated by the situation in the quantum Hall bilayer (QHB) at total filling factor one, i.e.
 with each layer  with half-filled lowest LL. In this system $p$-wave pairing between two kinds of non-relativistic Halperin-Lee-Read (HLR) composite fermions at intermediate interlayer distances was proposed in Ref. [\onlinecite{msr}], and, recently, this scenario was further substantiated by a detection of the topological signatures of the $p$-wave system in the torus geometry [\onlinecite{mdp}]. Therefore, it is natural to ask how this picture may be modified if we take into account the description by two Dirac CF's of the two half-filled LL  monolayers, and consider their possible pairing.

One of the main conclusions that we can draw by applying the Dirac CF formalism in the context of BCS pairing is that
due to the Dirac two-component nature, isotropic (gapped) pairing states may coexist with anisotropic ones, and this is in the accordance with the results on PH symmetric, single and bilayer fractional quantum Hall systems obtained by employing  exact diagonalization [\onlinecite{park,wy}], as well as  with experimental findings [\onlinecite{l1,l2}], in which anisotropy is probed by in-plane magnetic field. This may be a direct consequence of the dipole nature of CF's that is captured by Dirac formalism. Anisotropic pairing states may serve as seed states for Pfaffian and anti-Pfaffian through a process in which  PH asymmetry increases by introducing a mass term, while rotational symmetry gradually sets in. Furthermore, we find that the features, in particular low-energy spectrum, of the QHB at intermediate distances between the layers are better captured if we assume Dirac rather than HLR $p$-wave paired CF's at large distances (decoupled layers).
Already at the  effective field theory level, modeling the evolution with the distance between layers by Dirac CFs, we can detect the main feature of CF-composite boson (CB) mixed states [\onlinecite{msr,srm}]:  the decrease in the number of CF's with decreasing distance.

The paper is organized as follows. In Section II , based on Dirac formalism, we discuss
the single layer case and its pairing instabilities,
including the situation when the PH symmetry is spoiled by a mass term. In Section III we discuss the pairing instabilities in the bilayer system when the PH symmetry inside each layer is intact. In Section IV we examine the evolution of low-energy properties of the QHB with distance between layers, by including a mass term with an opposite sign in the two layers. The last section, Section V, is devoted to discussion and conclusions. Mean-field analysis of the coexistence of the isotropic and anisotropic pairings is presented in the Appendices.
% ------------------------------------
\section{Dirac composite fermion and Cooper pairing}
% ------------------------------------
We begin by considering a single Dirac fermion which was  proposed to effectively describe
half-filled lowest Landau level of electrons [\onlinecite{son}], with $s$-wave pairing between spinor components. The $s$-wave pairing suggested in Ref. [\onlinecite{son}], can be expressed by the following Bogoliubov - de Gennes Hamiltonian in the Nambu - Gorkov notation,
\begin{eqnarray}\label{BdGsD}
H &=&\frac{1}{2} \sum_{\bf k} \left[
  \begin{array}{cc}
   \Psi^\dagger ({\bf k}) & \tilde{\Psi}(-{\bf k}) \\
  \end{array}
\right]  \\ \nonumber
&\times&\left[
  \begin{array}{cc}
   {\cal D} ({\bf k}) & {\cal P}({\bf k}) \\
   {\cal P}^\dagger ({\bf k}) & - {\cal D} (-{\bf k}) \\
  \end{array}
\right]
\left[
  \begin{array}{c}
   \Psi ({\bf k}) \\
   \tilde{\Psi}^\dagger (-{\bf k}) \\
  \end{array}
\right],
\end{eqnarray}
where $\Psi ({\bf k})$ denotes a two-component spinor with momentum ${\bf k}$,
\begin{equation}
\Psi ({\bf k}) = \left[
  \begin{array}{c}
   \Psi_{a} ({\bf k}) \\
   \Psi_{b} ({\bf k}) \\
  \end{array}
\right], \,\,\,\,
\tilde{\Psi} ({\bf k}) = \left[
  \begin{array}{c}
   \Psi_{b} ({\bf k}) \\
   \Psi_{a} ({\bf k}) \\
  \end{array}
\right],
\end{equation}
and
\begin{equation}
{\cal D}({\bf k}) = \left[
  \begin{array}{cc}
   - \mu & k_x - i k_y \\
   k_x + i k_y & - \mu \\
  \end{array}
\right]=-\mu\sigma_0+k_x \sigma_x+k_y \sigma_y, \label{dmatrix}
\end{equation}
and  $2 \times 2$ matrix ${\cal P}({\bf k})$ describes Cooper pairing between $a$ and $b$ spinor components
\begin{equation}
{\cal P}({\bf k}) = \left[
  \begin{array}{cc}
   \Delta_s & 0\\
    0 & - \Delta_s \\
  \end{array}
\right]=\Delta_s \sigma_z  , \label{right_pairing_sD}
\end{equation}
or more explicitly
\begin{equation}
\delta {\cal H} = \sum_{\bf k} \{ - \Delta_s \Psi_{a}({\bf k}) \Psi_{b}(-{\bf k}) +  h.c. \}. \label{rightH}
\end{equation}
Here, $\sigma_0$ is the $2\times2$ identity matrix, while ${\bm \sigma}$ are the standard Pauli matrices. Throughout the paper we set $\hbar = 1$, and the Fermi velocity, $v_F = 1$.  $\mu$ denotes a chemical potential equal to $\mu = \sqrt{B} = k_F$, where $B$ and $k_F$ are the external magnetic field and Fermi
vector, respectively.

Since the pairing matrix anticommutes with the free Dirac Hamiltonian at the zero chemical potential, the dispersion of  Bogoliubons has
 the rotationally symmetric form
\begin{equation}
E_k^2 = (k \pm \mu)^2 + \Delta_{s}^2,
\end{equation}
where $k\equiv |{\bf k}|$.
This construction is considered in the literature
as a basis for a PH symmetric Pfaffian system.

However,  a different type of pairing is also possible with the pairing matrix
\begin{equation}
{\cal P}({\bf k}) = \left[
  \begin{array}{cc}
   0  & \alpha k_x + \beta k_y\\
    \alpha k_x - \beta k_y & 0 \\
  \end{array}
\right], 
\end{equation}
or more explicitly
\begin{eqnarray}
\delta {\cal H}'& =&\sum_{{\bf k}} \alpha k_x  \{ \Psi_{a}^\dagger ({\bf k}) \Psi_{a}^\dagger(-{\bf k}) + \Psi_{b}^\dagger({\bf k}) \Psi_{b}^\dagger(-{\bf k}) \} + h.c. \nonumber\\
 &+&\sum_{\bf k}  \beta k_y  \{ \Psi_{a}^\dagger({\bf k}) \Psi_{a}^\dagger(-{\bf k}) - \Psi_{b}^\dagger({\bf k}) \Psi_{b}^\dagger(-{\bf k}) \} + h.c.  \nonumber \\
\label{anisotropicH}
\end{eqnarray}
where $\alpha$ and $\beta$ are, in general, allowed to be complex coefficients. The overall form of $ \delta {\cal H}'$ is fixed by the requirement of the $CP$ symmetry, which, as emphasized in Ref. [\onlinecite{son}], is equivalent to the requirement of the PH symmetry in the real electron system. Namely, the $CP$ symmetry is a product of the charge conjugation, $C$,
\begin{equation}
C \Psi({\bf k}) C^{-1} = \sigma_x \Psi^* ({\bf k}),
\end{equation}
and a parity transformation, $P$,
\begin{equation}
P \Psi({\bf k}) P^{-1} =  \Psi^* ({\bf k}'),
\end{equation}
where ${\bf k} = (k_x, k_y) \rightarrow {\bf k}' = (k_x, - k_y)$ under  the parity transformation. Thus,
\begin{equation}
CP \Psi({\bf k}) (CP)^{-1} = \sigma_x \Psi({\bf k}').
\label{CP}
\end{equation}
The starting Dirac Hamiltonian\ (\ref{BdGsD}) with ${\cal P}=0$, as well  $ \delta {\cal H}'$ are both invariant under the $CP$ transformation  (\ref{CP}).  Notice that (\ref{rightH}) is invariant up to a sign change under $CP$ transformation. This is also a property of the small mass term that seems necessary to ensure the gauge invariance of the theory, and to avoid the singularity at ${\bf k} = 0$ [\onlinecite{ws2}]. The BCS pairing terms as the one in (\ref{rightH}) may accommodate the sign change by gauge transformations [\onlinecite{son}]. Thus the theory is invariant under $CP$ transformation in a more general sense, allowing for terms that are invariant up to a change of the sign. This
makes our choice for $p$ wave not unique. Indeed, other $p$ wave pairing order parameters are also possible, including one analogous to the A phase of ${^3}$He system that features two (gapless) Fermi points. This case can be analyzed analogously to the one considered here, and the main conclusions remain.  In the following, we restrict our discussion to the
$p$-wave case (\ref{anisotropicH}) invariant under $CP$ transformation in the strict sense.

We now consider the pairings given by Eq.\ (\ref{anisotropicH}), recently also discussed in Ref. [\onlinecite{wc}], in light of the possibility of introducing an anisotropy. The choice $\alpha = \Delta$ and  $\beta  = - i \Delta $ yields the pairing matrix, $ {\cal P}({\bf k})$, proportional to the Dirac Hamiltonian, ${\cal D}({\bf k})$, at chemical potential $\mu = 0$, and thus explicitly rotationally invariant. (See
 also Sec. III for further analysis of the rotational symmetry.). In that case, the dispersion relation of Bogoliubons, $E_{\bf k}^2 = k^2(1+\Delta^2)+\mu^2 \pm 2k \sqrt{\mu^2 + k^2 \Delta^2}$, implies that the the pairing just renormalizes
 chemical potential. On the other hand, by
 choosing $\alpha = \Delta$ and  $\beta  = + i \Delta $, we obtain
 \begin{equation}
E_{\bf k}^2 = k^2 (1 + \Delta^2) + \mu^2 \pm 2 \sqrt{\mu^2 k^2 + \Delta^2 (k_x^2 - k_y^2)^2}. \label{andispersion}
\end{equation}
This dispersion describes an {\em anisotropic} gapless system with four nodes at
\begin{equation}
k_x = \pm \frac{\mu}{\sqrt{1 - \Delta^2}}, \;\;\;\; {\rm and}\;\;\; k_y = 0,
\end{equation}
and
\begin{equation}
k_y = \pm \frac{\mu}{\sqrt{1 - \Delta^2}}, \;\;\;\; {\rm and}\;\;\; k_x = 0.
\end{equation}
The appearance of the four nodes related by the discrete $C_4$ symmetry is a consequence of the  $C_4$ symmetry of the pairing\ (\ref{anisotropicH}) with $\alpha = \Delta$ and  $\beta  = + i \Delta $.
In fact, Eq. (\ref{anisotropicH}) describes a whole family of gapless anisotropic solutions.

 If we consider both $s$-wave (\ref{rightH}) and $p$-wave (\ref{anisotropicH}) with $\alpha = \Delta$ and  $\beta  = + i \Delta $ pairings, the dispersion of the Bogoliubov quasiparticles is
\begin{equation}
\tilde{E}_{\bf k}^2 = \Delta_s^2 + k^2 (1 + \Delta^2) + \mu^2 \pm 2 \sqrt{\mu^2 k^2 + \Delta^2 (k_x^2 - k_y^2)^2},
\end{equation}
i.e. the dispersion (\ref{andispersion}) simply acquired a shift of
$\Delta_s^2$ in the presence of the $s$-wave pairing. This is a consequence of the anticommutation of the matrices corresponding to the two pairings, similarly to the situation in
Refs. [\onlinecite{RJ2014}, \onlinecite{roy-goswami}], which makes their coexistence likely at a finite chemical potential. Assuming a generic form of the two couplings driving the instabilities in the isotropic and anisotropic channels, in the presence of a small mass term, we show in Appendix A that the low energy description implies that the isotropic instability (\ref{rightH}) may coexist with the anisotropic one. This is consistent with experimental [\onlinecite{l1,l2}], and theoretical  [\onlinecite{park,wy}] findings pointing out that gapped states at half-filled Landau level can sustain and even harbor anisotropy.

In connection with the possible pairings given by Eq.\ (\ref{anisotropicH}) when $\alpha = \Delta$ and  $\beta  = + i \; \Delta $, we may notice that if we break $CP$ (particle-hole symmetry) by a mass term of the form $\sim \Psi^\dagger({\bf k}) \sigma_3 \Psi({\bf k})$,
one component, $a$ or $b$, of the Dirac field will remain in the low energy  sector. The remaining fermion should correspond to HLR (spinless) fermion which in turn pairs in the manner of $p$-wave. This should correspond to Pfaffian and anti-Pfaffian states (that comprise possible $( k_x \pm i k_y)$ states), in the absence of PH symmetry, but with an emergent rotational symmetry. A closely related proposal for the existence of the Pfaffian (Moore-Read) state in the presence of an excitonic instability already appeared in the context of Dirac CF physics in graphene
Ref. [\onlinecite{cai}].

 To further understand the pairings in Eqs.\ (\ref{rightH}) and (\ref{anisotropicH}), we now consider the chirality operator $\frac{{\bm\sigma} \cdot{\bf k}}{|k|}$, and its eigenstates
\begin{equation}
|+ \rangle = \frac{1}{\sqrt{2}} \left[
  \begin{array}{c}
   1 \\
   \frac{k_+}{k} \\
  \end{array}
\right], \,\,\,\,
|- \rangle = \frac{1}{\sqrt{2}} \left[
  \begin{array}{c}
   - 1 \\
   \frac{k_+}{k} \\
  \end{array}
\right].\label{chiral-eigenstates}
\end{equation}
We can  introduce Dirac operators with a definite chirality
\begin{equation}
\Psi_+ ({\bf k}) = \frac{1}{\sqrt{2}}( \Psi_a ({\bf k}) +  \frac{k_-}{k} \Psi_b ({\bf k})) ,
\end{equation}
and
\begin{equation}
\Psi_- ({\bf k}) = \frac{1}{\sqrt{2}}( - \Psi_a ({\bf k}) +  \frac{k_-}{k} \Psi_b ({\bf k})) ,
\end{equation}
to find that
\begin{equation}
\Psi_a ({\bf k})  \Psi_b ( - {\bf k})
=- \frac{1}{2}  \frac{k_+}{k} [\Psi_+ ({\bf k})  \Psi_+ ( - {\bf k}) +   \Psi_- ({\bf k}) \Psi_- (-{\bf k})],
\label{s_into_p}
\end{equation}
with $k_\pm\equiv k_x\pm i k_y$.
We can clearly see from Eq.\ (\ref{s_into_p}) that in the chirality basis i.e. the eigenbasis of the non-interacting system, the pairing (\ref{rightH}), in fact, describes a pairing in the odd ($p$-wave) channel. This
can be understood as a consequence of the non-trivial Berry phase contributions, as discussed in Ref. [\onlinecite{son}]; see also Ref. [\onlinecite{mn}]
for the influence of the singularities (topological charges) on the vorticity of Cooper pairs. On the other hand,
the anisotropic pairing (\ref{anisotropicH}) is a combination of odd channel components in the chirality basis.

We now analyze an alternative scenario for the coexistence with the $p$-wave pairing represented by the pairing matrix ${\cal P}({\bf k})=(\alpha k_x+\beta k_y)\sigma_x$ that features two Fermi points and does not require a mass for the coexistence with the isotropic state. In particular, as shown in Appendix B,  a special anisotropic pairing with
\begin{equation}\label{anisotropicHH}
{\cal P} \sim i  k_y\sigma_x
\end{equation}
can coexist with the isotropic pairing. Analogously, we can discuss pairing with ${\cal P}({\bf k})\sim(\gamma k_x+\delta k_y)\sigma_y$,
where $\gamma$ and $\delta$ are, in general, allowed to be complex coefficients.
The ensuing pairing is then given by
\begin{equation}\label{anisotropicHH2}
{\cal P}({\bf k})\sim k_x \sigma_y.
\end{equation}
Both these pairings
are invariant up to a change of sign (up to a gauge transformation) under the $CP$ transformation. Each pairing on its own features two Fermi points, and is likely energetically advantageous over the pairing in (\ref{anisotropicH}) that has four Fermi points. As we explicitly show in Appendix B, these pairings do not need a mass term to coexist with the isotropic state. Furthermore, in the presence of a mass term, they develop new components, and may thus evolve into the rotationally symmetric pairings of HLR fermions. These are the reasons that make states given by Eqs.\ (\ref{anisotropicHH}) or (\ref{anisotropicHH2}) likely present when considering pairing instabilities  in the half-filled LL, consistent with the exact diagonalization results  of Refs. [\onlinecite{park,wy}].

 Finally, we  point out that the Dirac based microscopic wave functions of pairing instabilities have not been proposed and
tested yet. The effective field theory approach seems currently to be the most efficient tool for treating the Dirac-based pairing instabilities and their properties. Once the microscopic description is provided, most importantly for the case of PH Pfaffian, anisotropic modifications may be induced in the manner described and discussed in Refs. [\onlinecite{h}, \onlinecite{bj2}].

% ----------------------------------
\section{Dirac fermions and $p$-wave  pairing}
% ------------------------------------

We consider the following general form of the Bogoliubov-de Gennes Hamiltonian,  motivated by the situation in a QHB system with each of the two layers at half filling,
\begin{eqnarray}\label{BdG}
H &=& \sum_{\bf k} \left[
  \begin{array}{cc}
   \Psi_\uparrow^\dagger ({\bf k}) & \Psi_\downarrow (-{\bf k}) \\
  \end{array}
\right]  \\ \nonumber
&\times&\left[
  \begin{array}{cc}
   {\cal D}_\uparrow ({\bf k}) & {\cal P}({\bf k}) \\
   {\cal P}^\dagger ({\bf k}) & - {\cal D}_\downarrow (-{\bf k}) \\
  \end{array}
\right]
\left[
  \begin{array}{c}
   \Psi_\uparrow ({\bf k}) \\
   \Psi_\downarrow^\dagger (-{\bf k}) \\
  \end{array}
\right],
\end{eqnarray}
where $\Psi_\uparrow ({\bf k})$ and $\Psi_\downarrow ({\bf k})$ are two component spinors,
\begin{equation}
\Psi_\uparrow ({\bf k}) = \left[
  \begin{array}{c}
   \Psi_{a \uparrow} ({\bf k}) \\
   \Psi_{b \uparrow} ({\bf k}) \\
  \end{array}
\right], \,\,\,\,
\Psi_\downarrow ({\bf k}) = \left[
  \begin{array}{c}
   \Psi_{b \downarrow} ({\bf k}) \\
   \Psi_{a \downarrow} ({\bf k}) \\
  \end{array}
\right].
\end{equation}
Matrices ${\cal D}_\uparrow ({\bf k})$ and ${\cal D}_\downarrow ({\bf k})$ describe two identical Dirac systems, ${\cal D}_\uparrow ({\bf k})={\cal D}_\downarrow ({\bf k})={\cal D} ({\bf k})$, with ${\cal D} ({\bf k})$ given by Eq.\ (\ref{dmatrix}), while $2 \times 2$ matrix ${\cal P}({\bf k})$ describes Cooper pairing between the two systems $\uparrow$ and $\downarrow$.

A triplet $p$-wave pairing between the same spinor components can be expressed
as the following term in the Hamiltonian,
\begin{eqnarray}
\delta {\cal H} &=& \sum_{\bf k} \{ [ \Delta^*_{\bf k} \Psi_{a \downarrow}(-{\bf k}) \Psi_{a \uparrow}({\bf k})  \\ \nonumber
&+&\Delta^*_{\bf k} \Psi_{b \downarrow}(-{\bf k}) \Psi_{b \uparrow}({\bf k}) ] \, + \, h.c. \},
\end{eqnarray}
with a pairing function $ \Delta_{\bf k} = \Delta (k_x \pm i k_y)$.
The corresponding pairing matrix in the Hamiltonian (\ref{BdG}) is
\begin{equation}
{\cal P}({\bf k}) = \left[
  \begin{array}{cc}
   0 & \Delta_{\bf k} \\
   \Delta_{\bf k} & 0 \\
  \end{array}
\right] = \Delta_{\bf k} \sigma_x.\label{anisotropic_pairing}
\end{equation}
A rotation around $z$ axis by an angle $\phi$ in both subsystems $\uparrow$ and $\downarrow$ is represented  by a matrix $ R = \exp (i \sigma_z  \phi/2)$ so that
\begin{eqnarray}
R \sigma_x R^{-1} &=& \sigma_x \cos \phi - \sigma_y \sin \phi, \\ \nonumber
R \sigma_y R^{-1} &=& \sigma_x \sin \phi + \sigma_y \cos \phi.
\end{eqnarray}
It can be readily seen that $ \tilde{R} H({\bf k}) \tilde{R}^{-1} \neq H({\bf k}')$ where
$ k_{x}^{'} = k_x\cos\phi  - k_y\sin\phi $ and $ k_{y}^{'} = k_x\sin\phi  + k_y\cos\phi $, and ${\tilde R}=\tau_0\otimes R$, with $\tau_0$ as the $2\times2$ unity matrix in the subsystem space.  Therefore, the system with the pairing matrix ${\cal P}({\bf k}) = \Delta_{\bf k}  \sigma_x$ is not rotationally invariant, and may lead to anisotropic behavior. In fact, the system is gapless, and supports two anisotropic Dirac cones at $ k_x^2 = \mu^2/(1 - \Delta^2) = k_0^2$ and $k_y = 0$. Expanding around $\pm k_0$ we obtain for $\Delta \ll 1$, $ E^2 \approx (1 - 2 \Delta^2) (\delta k_x)^2 + \Delta^2 (\delta k_y)^2$.
We find similar results if we choose,
\begin{equation}
{\cal P}({\bf k}) = \left[
  \begin{array}{cc}
   0 & \Delta_{\bf k} \\
   -\Delta_{\bf k} & 0 \\
  \end{array}
\right].\label{sign_anisotropic_pairing}
\end{equation}
Therefore the systems that we considered by now do not possess quantum spin Hall effect, and due to anisotropy are likely to be fragile under disorder, and certainly can not represent stable phases in realistic circumstances.

On the other hand, the system with the pairing matrix
\begin{equation}
{\cal P}({\bf k}) = \left[
  \begin{array}{cc}
   \Delta_{\bf k} & 0\\
    0 & - \Delta_{\bf k} \\
  \end{array}
\right] = \Delta_{\bf k} \sigma_z , \label{right_pairing}
\end{equation}
yields the dispersion relation of the Bogoliubons
\begin{equation}
E_{\pm} = \sqrt{(k \pm \mu)^2 + |\Delta_{\bf k}|^2}, \label{disper_triplet}
\end{equation}
and therefore resembles very closely $p$-wave pairing of ordinary fermions.
We now express the pairing in the chirality basis to obtain
\begin{eqnarray}
 &&\Delta^*_{\bf k} (\Psi_{b \downarrow}(-{\bf k}) \Psi_{a \uparrow}({\bf k}) -
 \Psi_{a \downarrow}(-{\bf k}) \Psi_{b \uparrow}({\bf k}))=- \Delta^*_{\bf k} \frac{1}{2}  \frac{k_+}{k} \nonumber \\
&\times&
 (\Psi_{+ \downarrow}(-{\bf k}) \Psi_{+ \uparrow}({\bf k}) -
 \Psi_{- \downarrow}(-{\bf k}) \Psi_{- \uparrow}({\bf k})). \nonumber \\
 \end{eqnarray}
 Thus depending whether  $ \Delta_{\bf k} = \Delta (k_x + i k_y)$ or
 $ \Delta_{\bf k} = \Delta (k_x - i k_y)$, we obtain $s$-wave or $d$-wave pairing,
 respectively, in the chirality basis.
 In this sense there is no surprise to find that the pairing matrix (\ref{right_pairing}) gives rise to a singlet state for $\uparrow$ and $\downarrow$ electrons. The choice for the pairing without the minus sign in Eq.\ (\ref{right_pairing}), i.e., ${\cal P}({\bf k})=\Delta_{\bf k}\sigma_0$, is not energetically favorable, since  pairing just renormalizes chemical potential in that case.

We now provide a topological characterization of pairing in Eq.\ (\ref{right_pairing}) through the  (pseudo)spin Chern number, $C_s$. In fact we find that in this case is $C_s = 1$, if we use the low-energy theory
with (\ref{right_pairing}) and $ \Delta_{\bf k} = \Delta (k_x + i k_y)$.  We calculated the Chern number by taking the eigenvectors
of the two lower Bogoliubov bands, $ |v_{-}({\bf k})\rangle$ and $ |v_{+}({\bf k})\rangle$,
corresponding to eigenvalues $ - E_{-}({\bf k})$ and $ - E_{+}({\bf k})$, respectively.
We first computed the Berry curvature of each vector,
\begin{equation}
F^{\sigma}_{xy}({\bf k}) = i  (\partial_x \langle v_{\sigma}({\bf k})| \partial_y |v_\sigma ({\bf k}) \rangle -  \partial_y \langle v_{\sigma}({\bf k})| \partial_x |v_\sigma ({\bf k}) \rangle ),
\end{equation}
and then the Chern number,
\begin{equation}
C_s = \frac{1}{2 \pi} \sum_\sigma \int d{\bf k} F^{\sigma}_{xy}({\bf k}),
\label{chern}
\end{equation}
where the sum in (\ref{chern}) is over the two lowest bands. Nevertheless, as discussed in the previous paragraph, and, also, due to the form of eigenvectors below, we expect that the real winding number is zero or two if a complete description is taken into account.

To further characterize  the pairing state, let us consider the four-component vectors of  the Bogoliubov bands with positive energy, $  E_{-}({\bf k})$ and $ E_{+}({\bf k})$,
\begin{eqnarray}
u_{-}(k) &=& \frac{1}{2 \sqrt{E_{-}}} \left\{ - \sqrt{E_{-} - (\mu - k)} \left(1, \frac{k_{+}}{k}\right),\right. \nonumber\\
 & & \left.\frac{\Delta \cdot k}{\sqrt{E_{-} - (\mu - k)}}  \left(-  \frac{k_{-}}{k},1\right) \right\},
\end{eqnarray}
and
\begin{eqnarray}
u_{+}(k) &=& \frac{1}{2 \sqrt{E_{+}}} \left\{  \sqrt{E_{+} - (\mu + k)} \left(1, - \frac{k_{+}}{k}\right),\right.\nonumber\\
 & &\left. \frac{\Delta \cdot k}{\sqrt{E_{+} - (\mu + k)}} \left(\frac{k_{-}}{k}, 1\right) \right\},
\end{eqnarray}
where we regrouped components to appear with common factors. In each  Bogoliubov eigenstate, the first two-component spinor, ( , ), is an eigenstate of the chirality
operator, given by Eq.\ (\ref{chiral-eigenstates}), while the second one is the eigenstate that is complex conjugated and with inverted components due to the ordering in the Nambu - Gorkov representation, and we fix $ \Delta_{\bf k} = \Delta (k_x + i k_y)$. From the coefficients in front of the fixed chirality states,  we find  the long-distance behavior of the pairing function ($g_{\bf k} \sim v_{\bf k}/u_{\bf k}$ in the usual BCS problem) in each band
\begin{equation}
g(z) \sim 1/|z| ,
\end{equation}
where $g(z)$ is the  pairing function in the real space, and $z = x + i y$. Thus the pairing function has the characteristic $s$-wave feature.

In this case the lowest gap
is at Fermi surface, $\Delta E \sim  \Delta \cdot k_F$, in  contrast with ordinary $p$-wave
pairing where the lowest gap is  at zero momentum, and it is equal to $\Delta E \sim k_F$ [\onlinecite{rg}].

% ------------------------------------
\section{Quantum Hall bilayer and $p$-wave paired composite fermions}
% ------------------------------------

 In light of recent advance in understanding  of each (isolated PH symmetric) half-filled  quantum Hall monolayer
based on Dirac CF's
it is quite natural to consider the physics of the bilayer, especially at the intermediate distances, in the same framework. It is important to take into account the $p$-wave
pairing [\onlinecite{msr}] which was initially expressed in terms of ordinary  HLR CF's. The picture based on the ordinary CF's does not have a clear answer for the lowest lying spectrum which appears nearly gapless (with small gap) or gapless when the system is put on a  torus, while the topological $p$-wave pairing of ordinary  fermions [\onlinecite{rg}] would likely produce a clear gap of the order $\mu$. However, even if we neglect possible insufficiencies with $p$-wave pairing of ordinary fermions, it is fundamentally important to address the problem of the QHB  in terms of Dirac CF's.

First we may  notice that the presence of the interlayer Coulomb interaction, which increases with decreasing distance between layers, spoils PH symmetry inside a layer.
We incorporate this breaking of the PH symmetry by introducing a mass $r$ in the Dirac matrices ${\cal D}_\uparrow ({\bf k})$ and ${\cal D}_\downarrow ({\bf k})$,  with opposite signs in each layer,
\begin{equation}
{\cal D}_\uparrow ({\bf k}) = \sigma_x k_x + \sigma_y k_y - \mu + r \sigma_z  = {\cal D}_\downarrow ({\bf k}).
\end{equation}
Second, the components of the  spinors in different layers are inverted with respect to each other, and thus the mass term  of the opposite sign in the two layers enters with the same sign in the matrices ${\cal D}_\uparrow ({\bf k})$ and ${\cal D}_\downarrow ({\bf k})$.
The dispersion relation in this case acquires the form
\begin{equation}
E_{\pm} = \sqrt{(\sqrt{k^2 + r^2} \pm \mu)^2 + |\Delta_{\bf k}|^2}. \label{disper_triplet_with_r}
\end{equation}
The masses in the two layers are of the opposite sign, due to the requirement of the PH symmetry of the whole system. Namely, under the transformation in each layer masses change sign [\onlinecite{son}], and if we, in addition, exchange layer index we recover the original Hamiltonian.

There are two important things to notice regarding the evolution of the CF state with increasing mass $r$:

(a) The minimum of the lower Bogoliubov band shifts from a finite value at $k_F^2 = \mu^2/(1 + \Delta^2)^2 - r^2$ to $k = 0$,
and this transition - without  closing of the gap - occurs at $r = \mu/(1 + \Delta^2)$;

  (b) Because   $k_F^2 = \mu^2/(1 + \Delta^2)^2 - r^2$, the Fermi momentum decreases with the mass, and therefore  the number of CF's reduces as the distance between the layers decreases.

Therefore the most important consequence of the assumed Dirac description of individual layers at large distances is that the number of CF's decreases as the distance between the layers decreases. For large distances we may assume that the pairing is weak, the order parameter
is small, and the pairing  cannot be detected then due to finite temperature effects, for instance. In any case we may choose, $ \Delta_{\bf k} = \Delta (k_x + i k_y)$, so that there is no Hall drag (pseudospin Hall effect) at large distances, but it develops gradually as the interlayer distance decreases and reaches  the  quantized value in agreement with experiments [\onlinecite{kim}]. This choice of the order parameter agrees with Refs. [\onlinecite{mdp,msr}].  For smaller distances $(r \sim \mu$ but $ r < \mu)$ we may assume that
the upper Bogoliubov band is pushed to high energies and an effective description in terms of quadratically dispersing CF's paired via weak $p$-wave pairing emerges,  implying an algebraically decaying Cooper pair wave function [\onlinecite{rg}].
The description of the system within this scenario then implies that at intermediate distances CB-CF mixture accounts for the total number of electrons [\onlinecite{srm}, \onlinecite{msr}]. As a consequence, composite bosons cannot have long range order, and likely have critical, algebraic pairwise correlations [\onlinecite{mdp}].

If at intermediate distances solely a collection of $p$-wave paired composite fermions, quadratically dispersing as in Ref. [\onlinecite{rg}], were present, signals of a topological phase with a large gap, $\Delta E \sim \mu$ would appear.
Instead, as detected on a torus in Ref. [\onlinecite{mdp}], there is an abundance of various low-energy excitations. This is in accordance with the above physical picture that implies a small portion of CF's at intermediate distances  in a topological phase with a small gap $\Delta E \sim \mu - r$, and $\mu \simeq r$.

As in the single layer case, anisotropic gapless solution\ (\ref{anisotropic_pairing}) is possible also for a bilayer. In the presence of a mass term $\sim r$ and in the case of the pairing  (\ref{anisotropic_pairing}) we obtain two anisotropic  Dirac cones at $ k_x^2 = \mu^2/(1 - \Delta^2) = k_0^2$ and $k_y = 0$. Expanding around $\pm k_0$ with $r \ll \mu$ we obtain  $ E^2 \approx (1 - 2 \Delta^2 - \frac{r^2}{\mu^2}) (\delta k_x)^2 + \Delta^2 (\delta k_y)^2$. The absence of a gap suggests a non-topological behavior. On the other hand, topological signatures were detected at intermediate distances in Ref. [\onlinecite{mdp}],
  in agreement with the characterization of isotropic weak $p$-wave pairing.  Thus the presence of the isotropic pairing, which may be accompanied by anisotropic ones,  seems crucial for the explanation of the properties at intermediate distances.

% ------------------------------------
\section{Discussion and Conclusions}
% ------------------------------------
The existence of anisotropic candidates for BCS paired states, in the case of monolayer (Sec. II), and bilayer (Sec. III and IV), is in agreement with the results in Ref. [\onlinecite{park}]. In that paper, the physics of the PH symmetric case of half-filled second Landau level is studied by exact diagonalization on a torus.  The main
 result of this numerical study is that the paired quantum Hall state in that case, as well closely related (by antisymmetrization) bilayer state, made of two kinds of electrons that each occupy quarter of the available single particle states in the second Landau level, are susceptible to anisotropic instabilities. By using the Dirac description of  the dipole nature of CFs, we can identify the paired quantum Hall state of Ref. [\onlinecite{park}] with PH Pfaffian, and its closeness to anisotropy as a sign of the relevance of  anisotropic solutions discussed in Sec. II. On the other hand, the relevance of the anisotropy for the bilayer state at effective $\nu=1/2 = 1/4 + 1/4$ total filling factor [\onlinecite{park}], may be again due to the composite - dipole nature of the CF's at filling factor $\nu=1/4$.
The Dirac description could be the easiest way to capture the dipole nature of a CF, despite the doubling of the fermionic degrees of freedom. In other words, we need particles and holes to describe dipoles [\onlinecite{sm}],  and the Dirac formalism could be a way to achieve that even in the cases when CF's have a Berry phase equal to $\pi/2$ (at quarter filling), with appropriate mass and chemical potential. If the Diracness is the cause of the anisotropic behavior, we can conclude that the Dirac formalism is equally applicable at $\nu=1/2$ and $\nu=1/4$. In this sense ``nothing is special at $\nu=1/2$" (Ref. [\onlinecite{fdmh}]) since only PH symmetry singles out Dirac description. The PH symmetry is sufficient but not necessary to cause the Diracness at the filling equal to one half.

If we restrict our discussion only to the case when CF's possess Berry phase equal to $\pi$, and thus Dirac formalism
seems appropriate for the bilayer case at total filling one, we demonstrated that the description by Dirac fermions is justified due to a global appearance and characterization of low-energy spectrum from the exact diagonalization on a torus [\onlinecite{mdp}]. In fact, the Dirac CF in the bilayer changes its Berry phase from value $\pi$ at large distances, to value $\sim 0$, at small distances (HLR fermion), while retaining its fermionic character. The second important consequence, due to the use of the Dirac formalism, is that the number of CF's is decreasing with the decreasing distance between layers. This is in in agreement with the necessity to use CF-CB mixed states to describe the bilayer at intermediate distances [\onlinecite{msr}].

Thus we can conclude that Dirac formalism can capture the basic phenomenology  of the bilayer at $\nu=1$, and the nature of the gapped paired states
in the single layer quantum Hall systems  with half-filled LL. We therefore expect it to become
an indispensable tool for further understanding of the paired states  in this context.

{\it Note added:} While this manuscript was in the final stage of preparation, Ref. [\onlinecite{wc}] appeared. It is a study of
possible pairings, based on Dirac formalism, and their realization in the case of a single layer with the half-filled LL.
 Ref. [\onlinecite{wc}] considered
pairings in the low-energy subspace of Dirac spectrum in
the context of a specific pairing mechanism.  In our work the low-energy projection is in place after the consideration of the pairing  instabilities within the Dirac formalism.  In this way we are able to account for the anisotropic pairings, with the consequences consistent with
theoretical and experimental findings, as we already emphasized.
\begin{acknowledgments}
We would like to thank S. Simon for a discussion.
The work was supported by
the Ministry of Education, Science, and Technological
Development of the Republic of Serbia under projects
ON171017 and  ON171031.
\end{acknowledgments}
\appendix
\section{Coexistence of the $CP$ invariant $p$-wave and $s$-wave pairings: Mean-field analysis}

The lower Bogoliubov band of the quadratic Hamiltonian, Eqs. (\ref{BdGsD}-\ref{right_pairing_sD}) with the additional pairing in Eq.\  (\ref{anisotropicH}) with $\alpha = \Delta$ and  $\beta  = + i \Delta $,  and a mass term $r \Psi^\dagger({\bf k}) \sigma_3 \Psi({\bf k})$,
 is
\begin{eqnarray}
E^2&=&\mu^2 + \Delta_s^2 + r^2 + k^2 + \Delta^2 k^2  \nonumber \\
&-&  2 \sqrt{\mu^2 (k^2 + r^2) + [\Delta (k_x^2 - k_y^2) + \Delta_s r]^2 }. \nonumber
\end{eqnarray}

We analyze the pairing instabilities in the low-energy theory by introducing a cut-off $\Lambda$, so that relevant momenta from the interval around Fermi energy are defined by $\Lambda$,  $ k\in (\mu - \Lambda , \mu + \Lambda)$. Also we assume that $ \mu \Delta \ll \Delta_s \ll \Lambda \ll \mu$, and, at zero temperature, estimate the free energy when both isotropic $(\Delta_s)$ and anisotropic $(\Delta)$ pairings  are present. From the BCS mean field decoupling of effective attractive interactions we obtain terms proportional to the order parameters
$\Delta^2$  and $\Delta_s^2$ (condensate energy) besides the contribution arising from the quasiparticles in the lower Bogoliubov band. (The upper band is assumed effectively a constant due to the constraint on the momenta.) The free energy density, ${\cal F}/{A}$, then reads
\begin{eqnarray}
\frac{{\cal F}}{A} & = & g_1 \Delta_s^2 + g_2 \Delta^2 - \frac{\mu}{4 \pi} \Lambda^2 \nonumber \\
&-&   \frac{\mu}{4 \pi} \{(1 + \ln \frac{4 \Lambda^2}{\Delta_s^2})  {\cal M}\} , \nonumber \\ \label{M0A}
\end{eqnarray}
where
\begin{equation}
{\cal M} = \Delta_s^2  + \frac{\Delta^2 \mu^2}{4} - \frac{r}{2} \Delta_s \Delta,  \label{MA}
\end{equation}
with $g_1$ and $g_2$ as positive coupling constants which drive the instabilities in the respective channels.
Here, we assume  $ r \ll \frac{\Delta_s}{\Lambda} ( \mu\Delta)$.

We derive Eq. (\ref{M0A}) with (\ref{MA}) by expanding the square root for  large $\mu$, and then performing the integral over $k$ (i.e.
radial component of vector ${\bf k}$). Before the final angular integration, we further simplified the result of the $k$
integration by assuming the stated ordering of scales.

In the BCS weak coupling limit,
by minimizing the free energy i.e. the total ground state energy, we obtain
\begin{eqnarray}
\Delta_s & \approx & 2 \Lambda \exp\{- \frac{2 \pi g_1}{\mu}\}, \nonumber \\
\Delta  & \approx &  \frac{r \tilde{g_1}}{\tilde{g_1} \mu^2 - 4 g_2} \Delta_s ,
\end{eqnarray}
where $ \tilde{g_1} = \frac{\mu}{4 \pi} + g_1$.
Thus we can conclude that for $ \mu \Delta \ll \Delta_s \ll \Lambda \ll \mu$, and in the presence of small mass $r$, the isotropic instability can be accompanied by   the anisotropic pairing.  This is due to the cross term in ${\cal F}$ with $\Delta_s$ and $\Delta$ - see Eqs.\ (\ref{MA}) and (\ref{M0A}). This may also be understood from the fact that the matrices corresponding to the isotropic and anisotropic pairings anticommute.

\section{Coexistence of the $CP$ asymmetric $p$-wave and $s$-wave pairings: Mean-field analysis}
 Here we discuss a pairing defined by
\begin{equation}
{\cal P}({\bf k}) = \left[
  \begin{array}{cc}
   0  & \alpha k_x + \beta k_y\\
    \alpha k_x + \beta k_y & 0 \\
  \end{array}
\right] = (\alpha k_x+\beta k_y)\sigma_x , \label{AppBsecond_pairing_sD}
\end{equation}
or in terms of the spinors, as a part of the complete Hamiltonian,
\begin{eqnarray}
 \sum_{\bf k}(\alpha k_x + \beta k_y) \{ \Psi_{a}({\bf k}) \Psi_{a}(-{\bf k}) + \Psi_{b}({\bf k}) \Psi_{b}(-{\bf k}) \} + h.c. , \nonumber \\
\label{anisotropicHA}
\end{eqnarray}
where $\alpha$ and $\beta$ are, in general, allowed to be complex coefficients.

The lower Bogoliubov band of the quadratic Hamiltonian, Eqs. (\ref{BdGsD}-\ref{right_pairing_sD}) with the additional pairing in (\ref{anisotropicHA}), is
\begin{eqnarray}
E^2&=&\mu^2 + \Delta_s^2 + k^2  (1 + f_1^2 + f_2^2) \nonumber \\
&-&  2k\sqrt{\mu^2  + \Delta_s^2 f_2^2 + k_x^2 (f_1^2 + f_2^2) - 2 \Delta_s f_2 \frac{k_y}{k} \mu }. \nonumber
\end{eqnarray}
Here, $\alpha k_x + \beta k_y = k (f_1 + i f_2)$ where $ f_i = \alpha_i \cos \phi + \beta_i \sin \phi, i =1,2$ and $\alpha_1 , \alpha_2, \beta_1,$ and $ \beta_2 $ are real, and $\phi$ is the polar angle of the momentum vector.

As in Appendix A, here we also analyze the pairing instabilities in the low-energy theory by introducing a cut-off $\Lambda$, so that relevant momenta from the interval around Fermi energy are defined by $\Lambda$,  $ k\in (\mu - \Lambda , \mu + \Lambda)$. Also we assume that $ \mu \omega \ll \Delta_s \ll \Lambda \ll \mu$, where $\omega$ can be $\alpha_1$, $\alpha_2, \beta_1,$ or $ \beta_2$, and, at zero temperature, estimate the free energy when both, isotropic $(\Delta_s)$ and anisotropic $(f_1 , f_2)$ pairings  are present. From the BCS mean field decoupling of effective attractive interactions we have terms proportional to
$f_1^2 , f_2^2$ (averaged over angles) and $\Delta_s^2$ next to the contribution from the lower Bogoliubov band. (The upper band is assumed effectively a constant due to the constraint on the momenta.) The free energy density, ${\cal F}/{A}$, then reads
\begin{eqnarray}
\frac{{\cal F}}{A} & = & g_1 \Delta_s^2 + g_2 (\alpha_1^2 + \alpha_2^2 + \beta_1^2 + \beta_2^2) \nonumber \\
&-&  \frac{1}{2} \frac{1}{(2\pi)^2} \{ \mu \; \Lambda^2 \; 2 \pi + (1 + \ln \frac{4 \Lambda^2}{\Delta_s^2}) \times \pi \mu {\cal M}\} \label{M0} \nonumber \\
\end{eqnarray}
where
\begin{equation}
{\cal M} = \Delta_s^2  + \frac{1}{2}
\Delta_s \beta_2 \mu + \frac{1}{4} (\alpha_1^2 + \alpha_2^2 + 3 \beta_1^2 + 3 \beta_2^2) \mu^2, \label{M}
\end{equation}
with $g_1$ and $g_2$ as positive coupling constants which drive the instabilities in the respective channels.
The last contribution of the quadratic order in anisotropic parameters, proportional to $ \sum_i (\alpha_i^2 + 3 \beta_i^2)$ was derived assuming $ \Lambda \ll \frac{\beta_2 \mu}{\Delta_s} \mu $.

To find this result for the free energy density we applied the same set of approximations as in Appendix A. We derived Eq. (\ref{M0}) with (\ref{M}) by expanding the value of the square root for  large $\mu$, and then performing the integral over $k$. Before the final angular integration, we further simplified the result of the integration over $k$
by assuming the stated ordering of scales.

In the BCS weak coupling limit,
by minimizing the free energy i.e. the total ground state energy, assuming $ \Delta_s \gg \omega \mu$ where $\omega$ can be $\alpha_1$, $\alpha_2, \beta_1,$ or $ \beta_2$, we obtain
\begin{eqnarray}
\Delta_s & \approx & 2 \Lambda \exp\{- \frac{4 \pi g_1}{\mu}\}, \nonumber \\
\beta_2  & \approx & \frac{\mu^2}{32 \pi} \frac{1}{g_2} \Delta_s (1 + \frac{8 \pi g_1}{\mu}), \nonumber \\
\alpha_1 & = & \alpha_2  = \beta_1 = 0.
\end{eqnarray}
Thus we can conclude that for cut-off $\Lambda$, $ \mu \beta_2 \ll \Delta_s \ll \Lambda \ll \mu$, and in the presence of the isotropic instability $\Delta_s$ we can expect  the presence of the anisotropic pairing with the order parameter $\sim i \beta_2 k_y$. This is due to the cross term in ${\cal F}$ with $\Delta_s$ and $\beta_2$ - see Eqs.\ (\ref{M}) and (\ref{M0}).

In the presence of mass $r$ the dispersion of the Bogoliubons is modified as
\begin{eqnarray}
E^2&=&\mu^2 + r^2 + \Delta_s^2 + k^2  (1 + f_1^2 + f_2^2) \nonumber \\
&-&  2 \sqrt{ \mu^2 k^2 + \Delta_s^2 f_2^2 k^2 + k_x^2 (f_1^2 + f_2^2)k^2 + {\cal R}}, \nonumber
\end{eqnarray}
where
\begin{eqnarray}
{\cal R}&=& r^2 (\Delta_s^2 + \mu^2) \nonumber \\
&+&  2 \Delta_s k (- k_y f_2 \mu + k_x f_1 r).
\end{eqnarray}
We can notice that besides the cross term $ \sim \Delta_s f_2 k_y$ under square root in the above equation, we have, in the presence of a mass $r$, the term $ \sim \Delta_s f_1 k_x$. By performing the similar mean field analysis as before, we can find that this term will lead to the development of the real component proportional to  $k_x$ in the anisotropic pairing, $\alpha k_x + \beta k_y = \alpha_1 k_x + i \beta_2 k_y$, with $ \alpha_1 /\beta_2 \sim r/\mu$ for $r \ll \mu$. Eventually , for $r \lesssim \mu$, we expect that $\Delta_s = 0$, and
the presence of the rotationally symmetric $p$ wave, $\alpha k_x + \beta k_y \sim (k_x \pm i  k_y)$ of one-component quadratically dispersing HLR composite fermions. Indeed the assumption $\Delta_s = 0$, and the presence of the $p$ wave
are compatible with $ r < \mu$, and there is no closing of the gap.

\end{document}